\documentclass[preprint,showpacs,preprintnumbers,amsmath,amssymb]{revtex4}
\usepackage{epsfig}
\usepackage{graphicx}
\usepackage{dcolumn}
\usepackage{bm}
\usepackage{threeparttable}

\def\beq{\begin{equation}}
\def\eeq{\end{equation}}
\def\eeqn{\end{equation}}
\newcommand\iden{\leavevmode\hbox{\small1\normalsize\kern-.33em1}}

\newcommand{\bea} {\begin{eqnarray}}
\newcommand{\eea} {\end{eqnarray}}

\let\jnfont=\rm
\def\NPB#1 {{\jnfont Nucl.\ Phys.\ B }{\bf #1} }
\def\PLB#1 {{\jnfont Phys.\ Lett.\ B }{\bf #1} }
\def\EPJC#1 {{\jnfont Eur.\ Phys.\ Jour.\ C }{\bf #1} }
\def\PRD#1 {{\jnfont Phys.\ Rev.\ D }{\bf #1} }
\def\PRL#1 {{\jnfont Phys.\ Rev.\ Lett.\ }{\bf #1} }
\def\MPLA#1 {{\jnfont Mod.\ Phys.\ Lett.\ A }{\bf #1} }
\def\JPG#1 {{\jnfont J.\ Phys.\ G }{\bf #1} }
\def\CTP#1 {{\jnfont Commun.\ Theor.\ Phys.\ }{\bf #1} }
\def\JHEP#1 {{\jnfont JHEP \ }{\bf #1} }
\def\NPPS#1 {{\jnfont Nucl.\ Phys.\ Proc.\ Suppl.\ }{\bf #1} }
\def\CPC#1 {{\jnfont Comput.\ Phys.\ Commun.\ }{\bf #1} }
\def\CPL#1 {{\jnfont Chin.\ Phys.\ Lett. }{\bf #1} }
\def\APPB#1 {{\jnfont Acta\ Phys.\ Polon.\ B }{\bf #1} }

\def\lsim{\raise0.3ex\hbox{$<$\kern-0.75em\raise-1.1ex\hbox{$\sim$}}}
\def\gsim{\raise0.3ex\hbox{$>$\kern-0.75em\raise-1.1ex\hbox{$\sim$}}}
\def\PR#1 {{\jnfont Phys.\ Rept. }{\bf #1} }
\def\CHC#1 {{\jnfont Chin.\ Phys.\ C }{\bf #1} }
\def\NIMA#1 {{\jnfont Nucl.\ Instrum.\ Meth.\ A }{\bf #1} }

\begin{document}

\title{\ \\[10mm]Probing a pseudoscalar at the LHC in light of $R(D^{(*)})$ and muon $g-2$ excesses}

\author{Lei Wang$^{1}$, Jin Min Yang$^{2,3}$, Yang Zhang$^{2}$}

\affiliation{$^1$ Department of Physics, Yantai University, Yantai 264005, P. R. China\\
$^2$ CAS Key Laboratory of Theoretical Physics,
     Institute of Theoretical Physics, Academia Sinica, Beijing 100190, P. R. China\\
$^3$ School of Physics, University of Chinese Academy of Sciences, Beijing 100049, P. R. China}


\begin{abstract}
We study the excesses of $R(D^{(*)})$ and muon $g-2$ in the framework
of a two-Higgs-doublet model
with top quark flavor-changing neutral-current (FCNC) couplings.
Considering the relevant theoretical
and experimental constraints, we find that the $R(D^{(*)})$ and muon $g-2$
excesses can be simultaneously explained in a parameter space allowed by
the constraints. In such a parameter space the pseudoscalar ($A$) has a mass between
20 GeV and 150 GeV so that it can be produced from the top quark FCNC decay $t\to A c$.
Focusing on its dominant decay $A\to \tau\bar{\tau}$,
we perform a detailed simulation on $pp\to t\bar{t}\to Wb Ac\to jjbc\tau\bar{\tau}$ and
find that the $2\sigma$ upper limits from a data
set of 30 (100) fb$^{-1}$ at the 13 TeV LHC can mostly (entirely) exclude such a parameter space.
\end{abstract}

\maketitle

\section{Introduction}
Recently, the BaBar \cite{babar1,babar2}, Belle \cite{belle1,belle2}
and LHCb \cite{lhcb} collaborations have reported anomalies in the
ratios \beq R(D^{(*)})=\frac{{\mathcal B}(\bar{B}\to
D^{(*)}\tau\bar{\nu}_\tau)}{{\mathcal B}(\bar{B}\to
D^{(*)}\ell\bar{\nu}_\ell)}, \eeq where $\ell=e,~\mu$. The average
values from the Heavy Flavor Average Group is \cite{average}
\begin{align}\label{eq:experiment}
R(D)_{\rm avg}=&0.397\pm0.040\pm0.028\,, \\
R(D^\ast)_{\rm avg}=&0.316\pm0.016\pm0.010\,.
\end{align}
Compared to the SM predictions
\begin{align}\label{eq:RD}
R(D)_{\rm SM}=&0.300\pm 0.008\,,\\
R(D^\ast)_{\rm SM}=&0.252 \pm 0.003\,,
\end{align}
there is a discrepancy of $1.9\sigma$ for $R(D)$ and $3.3\sigma$ for
$R(D^*)$. The anomalies have been studied in some specific new
physics models
\cite{Freytsis:2015qca,Calibbi:2015kma,Deppisch:2016qqd,Dumont:2016xpj,Dorsner:2016wpm,Sakaki:2014sea,Bauer:2015knc,
Fajfer:2015ycq,Sahoo:2015pzk,Barbieri:2015yvd,Sakaki:2013bfa,Boucenna:2016wpr,Hati:2015awg,Greljo:2015mma,
Das:2016vkr,Zhu:2016xdg,Deshpande:2012rr,160509308,rdbu1,rdbu2,rdbu3,rdbu4,rdbu5,rdbu6,rdbu7,rdbu8,rdbu9,rdbu10,rdbu11,rdbu12},
including the possibility of a charged Higgs boson
~\cite{rd1,rd2,rd3,12108443,Cline:2015lqp,150900491,Crivellin:2015hha,
Hwang:2015ica,Crivellin:2013wna,Sakaki:2012ft,Nierste:2008qe,Kiers:1997zt,Tanaka:1994ay,
Hou:1992sy}.

On the other hand, the muon anomalous magnetic moment ($g-2$) is a
very precisely measured observable. The muon $g-2$ anomaly has been
a long-standing puzzle since the announcement by the E821 experiment
in 2001 \cite{mug21,mug22}. There is an approximate $3\sigma$
discrepancy between the experimental value and the SM prediction
\cite{mu3sig1,mu3sig2,mu3sig3}. The muon $g-2$ anomaly can be simply
explained in the two-Higgs-doublet model (2HDM)
\cite{mu2h1,mu2h2,mu2h3,mu2h4,mu2h5,mu2h6,mu2h7,mu2h8,mu2h9,mu2h10,mu2h11,mu2h12,mu2h13,mu2h135,mu2h14,mu2h15,mu2h16}.

In this paper, we examine the $R(D^{(*)})$ and muon g-2
excesses in a 2HDM with the top quark flavor-changing
neutral-current (FCNC) couplings. We consider various theoretical
and experimental constraints from the precision electroweak data,
the $B$-meson decays, the $\tau$ decays as well as the observables of
the top quark and Higgs searches. In this model, the lepton Yukawa
couplings can simultaneously affect $R(D^{(*)})$, muon g-2 and
the lepton universality from $\tau$ decays, and thus these three
observables have a strong correlation. The $R(D^{(*)})$ and muon
g-2 excesses favor a light pseudoscalar with a large coupling to
lepton and nonzero top FCNC couplings, which implies that the
pseudoscalar can be produced from the top quark FCNC decay $t\to A c$
and then dominantly decays in the mode $A\to \tau\bar{\tau}$.
We will perform a detailed simulation on the signal $pp\to t\bar{t}\to Wb Ac\to jjbc\tau\bar{\tau}$
and the corresponding backgrounds at the LHC.

Our work is organized as follows. In Sec. II we recapitulate the
2HDM with the top quark FCNC couplings. In Sec. III we perform
numerical calculations. In Sec. IV, we discuss the $R(D^{(*)})$ and
muon g-2 excesses after imposing the relevant theoretical and
experimental constraints, and then perform the simulations on $pp\to
t\bar{t}\to Wb Ac\to jjbc\tau\bar{\tau}$. Finally, we give our
conclusion in Sec. V.

\section{two-Higgs-doublet model with top quark FCNC couplings}
The general Higgs potential is written as
\cite{2h-poten}
\begin{eqnarray} \label{V2HDM} \mathrm{V} &=& m_{11}^2
(\Phi_1^{\dagger} \Phi_1) + m_{22}^2 (\Phi_2^{\dagger}
\Phi_2) - \left[m_{12}^2 (\Phi_1^{\dagger} \Phi_2 + \rm h.c.)\right]\nonumber \\
&&+ \frac{\lambda_1}{2}  (\Phi_1^{\dagger} \Phi_1)^2 +
\frac{\lambda_2}{2} (\Phi_2^{\dagger} \Phi_2)^2 + \lambda_3
(\Phi_1^{\dagger} \Phi_1)(\Phi_2^{\dagger} \Phi_2) + \lambda_4
(\Phi_1^{\dagger}
\Phi_2)(\Phi_2^{\dagger} \Phi_1) \nonumber \\
&&+ \left[\frac{\lambda_5}{2} (\Phi_1^{\dagger} \Phi_2)^2 + \rm
h.c.\right]+ \left[\lambda_6 (\Phi_1^{\dagger} \Phi_1)
(\Phi_1^{\dagger} \Phi_2) + \rm h.c.\right] \nonumber \\
&& + \left[\lambda_7 (\Phi_2^{\dagger} \Phi_2) (\Phi_1^{\dagger}
\Phi_2) + \rm h.c.\right].
\end{eqnarray}
We focus on the CP-conserving model in which all $\lambda_i$ and
$m_{12}^2$ are real. We assume $\lambda_6$ and $\lambda_7$
are zero, and thus the Higgs potential has a softly broken $Z_2$ symmetry.
The two complex scalar doublets have the hypercharge $Y = 1$:
\begin{equation}
\Phi_1=\left(\begin{array}{c} \phi_1^+ \\
\frac{1}{\sqrt{2}}\,(v_1+\phi_1^0+ia_1)
\end{array}\right)\,, \ \ \
\Phi_2=\left(\begin{array}{c} \phi_2^+ \\
\frac{1}{\sqrt{2}}\,(v_2+\phi_2^0+ia_2)
\end{array}\right),
\end{equation}
where the electroweak vacuum expectation values (VEVs) $v^2 = v^2_1
+ v^2_2 = (246~\rm GeV)^2$, and the ratio of the two VEVs is defined
as $\tan\beta=v_2 /v_1$. After spontaneous electroweak
symmetry breaking, there are five mass eigenstates: two neutral
CP-even $h$ and $H$, one neutral pseudoscalar $A$, and two charged
scalars $H^{\pm}$.

The general Yukawa interactions in the physical basis are given as
 \bea
- {\cal L} &=&Y_{u1}\,\overline{Q}_L \, \tilde{{ \Phi}}_1 \,u_R
+\,Y_{d1}\,
\overline{Q}_L\,{\Phi}_1 \, d_R\, + \, Y_{\ell 1}\,\overline{L}_L \, {\Phi}_1\,e_R\nonumber\\
&&+Y_{u2}\,\overline{Q}_L \, \tilde{{ \Phi}}_2 \,u_R\,
+\,Y_{d2}\, \overline{Q}_L\,{\Phi}_2 \, d_R\,+\, Y_{\ell 2}
\overline{L}_L\, {\Phi}_2\,e_R \,+\, \mbox{h.c.}\,, \eea where
$Q_L^T=(u_L\,,d_L)$, $L_L^T=(\nu_L\,,l_L)$,
$\widetilde\Phi_{1,2}=i\tau_2 \Phi_{1,2}^*$, and $Y_{u1,2}$,
$Y_{d1,2}$ and $Y_{\ell 1,2}$ are $3 \times 3$ matrices in family
space.

To avoid the tree-level FCNC of the down-type quarks and leptons,
we take the two Higgs doublet fields to have the aligned
Yukawa coupling matrices \cite{a2hm1,a2hm2}:
\bea
&&Y_{\ell_1}=c_\ell~\rho_\ell,~~Y_{\ell_2}=s_\ell~ \rho_\ell,  \\
&&Y_{d1}=c_d~ \rho_d,~~Y_{d2}=s_d~ \rho_d, \eea where $c_d\equiv \cos\theta_d$,
$s_d\equiv \sin\theta_d$, $c_\ell\equiv
\cos\theta_\ell$, $s_\ell\equiv \sin\theta_\ell$,  and $\rho_d$ ($\rho_\ell$) is the $3 \times 3$
matrix.

For the Yukawa coupling matrices of the up-type quarks, we take
\bea
&&X_{ii}=\frac{\sqrt{2}m_{ui}}{v}(s_\beta+c_\beta
\kappa_u), \\
&&X_{ct}=\frac{\sqrt{2m_c m_t}}{v} c_\beta \lambda_{ct}, \\
&&X_{tc}=X_{ct},
\eea
where $X=V_L Y_{u2}
V_R^{\dagger}$, and  $V_L$ $(V_R)$ is the unitary matrix which
transforms the interaction eigenstates to the mass eigenstates for the
left-handed (right-handed) up-type quark fields.
We adopt the Cheng-Sher ansatz for $X_{ct}$ and $X_{tc}$ \cite{Cheng-Sher},
and other non-diagonal matrix elements of $X$ are taken as zero.

The Yukawa couplings of the neutral Higgs bosons are given as
\bea\label{hffcoupling} &&
y_{hf_if_i}=\frac{m_{f_i}}{v}\left[\sin(\beta-\alpha)+\cos(\beta-\alpha)\kappa_f\right], \nonumber\\
&&y_{Hf_if_i}=\frac{m_{f_i}}{v}\left[\cos(\beta-\alpha)-\sin(\beta-\alpha)\kappa_f\right], \nonumber\\
&&y_{Af_if_i}=-i\frac{m_{f_i}}{v}\kappa_f~{\rm (for~u)},~~~~y_{Af_if_i}=i \frac{m_{f_i}}{v}\kappa_f~{\rm (for~d,~\ell)},\nonumber\\
&&y_{hct}=\cos(\beta-\alpha)\frac{\sqrt{m_c m_t}}{v}\lambda_{ct},~~~~~~~~y_{htc}=y_{hct},\nonumber\\
&&y_{Hct}=-\sin(\beta-\alpha)\frac{\sqrt{m_c m_t}}{v}\lambda_{ct},~~~~y_{Htc}=y_{Hct},\nonumber\\
&&y_{Act}=-i\frac{\sqrt{m_c
m_t}}{v}\lambda_{ct},~~~~~~~~~~~~~~~~~~y_{Atc}=y_{Act}, \eea where
$\kappa_d\equiv-\tan(\beta-\theta_d)$ and
$\kappa_\ell\equiv-\tan(\beta-\theta_\ell)$.

The Yukawa interactions of the charged Higgs are given as
\begin{align} \label{eq:Yukawa2}
 \mathcal{L}_Y & = - \frac{\sqrt{2}}{v}\, H^+\, \Big\{\bar{u}_i \left[\kappa_d\,(V_{CKM})_{ij}~ m_{dj} P_R
 - \kappa_u\,m_{ui}~ (V_{CKM})_{ij} ~P_L\right] d_j + \kappa_\ell\,\bar{\nu} m_\ell P_R \ell
 \Big\}\nonumber\\
&~~~~ - \frac{\sqrt{2m_c m_t}}{v}\lambda_{ct}\, H^+\, \left[-\bar{u}_m
  ~ (V_{CKM})_{nj} ~P_L~d_j
\right]+h.c.,
 \end{align}
where $i,j=1,2,3$, and $m,n=2,3$ with $m\neq n$.

The neutral Higgs boson couplings with the gauge bosons normalized to the
SM are given by
\beq
y^h_{V}=\sin(\beta-\alpha),~~~
y^H_{V}=\cos(\beta-\alpha),\label{hvvcoupling}
\eeq
where $V$ denotes $Z$ or $W$.

The non-diagonal matrix elements $X_{ct}$ and $X_{tc}$ lead to the top quark FCNC of $h$, $H$ and $A$, and
give additional contributions to the couplings of charged Higgs and top quark (charm quark), as shown in
Eq. (\ref{hffcoupling}) and Eq. (\ref{eq:Yukawa2}). In the exact alignment limit \cite{alignment1,alignment2},
namely $\cos(\beta-\alpha)=0$, from Eq. (\ref{hffcoupling}) and Eq. (\ref{hvvcoupling}) we find that for $h$
the couplings to the fermions and gauge bosons are same as in the SM, and the tree-level top quark FCNC
couplings are absent. The heavy CP-even Higgs ($H$) has no couplings to the gauge bosons.

\section{Numerical calculations}
In our analysis we take the light CP-even Higgs boson $h$ as the
SM-like Higgs, $m_h=125$ GeV. In order to avoid the constraints from
the searches for the top quark FCNC of the SM-like Higgs, we take
the exact alignment limit, namely $\sin(\beta-\alpha)=1$. The muon
$g-2$ favors a light pseudoscalar with a large coupling to the
lepton, and a sizable mass splitting between $H$ and $A$. The
precision electroweak data favors a small mass splitting of $H$ and
$H^{\pm}$. Therefore, we take
\begin{eqnarray}
&&20 {\rm\  GeV} \leq ~m_{A}  \leq 180 {\rm\  GeV},~~~ -150\leq \kappa_\ell \leq -30,\nonumber\\
&&200 {\rm\  GeV} \leq ~m_{H}  \leq 700  {\rm\  GeV},~~200 {\rm\  GeV} \leq ~m_{H^\pm}  \leq 700  {\rm\  GeV}.
\end{eqnarray}

In order to loose the constraints from the searches for $pp\to
A~ (H)\to\tau \bar{\tau}$ and the constraints from observables of
down-type quarks, we take
\beq
\kappa_u=\kappa_d=-1/\kappa_\ell,
\eeq
which is similar to the Yukawa couplings of the lepton-specific
2HDM. For a very large $\kappa_\ell$, this choice is equivalent to
assume $\kappa_u$ and $\kappa_d$ to be negligible.

The other free parameters are randomly scanned in the following ranges
\beq
0<\lambda_{ct}<30,~-(400~{\rm GeV})^2 \leq m_{12}^2 \leq (400~{\rm GeV})^2,~0.1\leq\tan\beta\leq 10.
\eeq
Note that the $R(D^{(*)})$ excess favors opposite signs between $\lambda_{ct}$ and $\kappa_\ell$ \cite{150900491}.

In our scan, we consider the following observables and constraints:
\begin{itemize}

\item[(1)] Theoretical constraints and precision electroweak data. We
use $\textsf{2HDMC}$ \cite{2hc-1,2hc-2} to implement the theoretical
constraints from the vacuum stability, unitarity and
coupling-constant perturbativity, as well as the constraints from
the oblique parameters ($S$, $T$, $U$) and $\delta\rho$.

\item[(2)] The muon $g-2$. At the one-loop level, the muon $g-2$ is corrected by
\cite{mu2h1,mua1loop1,mua1loop2}
\beq
    \Delta a_{\mu1} =
    \frac{1}{8 \pi^2 } \, \sum_{\phi = h,~ H,~ A ,~ H^\pm}
    |y_{\phi\mu\mu}|^2  r_{\phi\mu} \, f_\phi(r_{\phi\mu}),
\label{amuoneloop}
\end{equation}
where $r_{\phi\mu} =  m_\mu^2/m_\phi^2$ and $y_{H^\pm\mu\mu}=y_{A\mu\mu}$.
For $r_{\phi\mu}\ll$ 1 we have
\beq
    f_{h,H}(r) \simeq- \ln r - 7/6,~~
    f_A (r) \simeq \ln r +11/6, ~~
    f_{H^\pm} (r) \simeq -1/6.
    \label{oneloopintegralsapprox3}
\eeq
The muon $g-2$ can be also corrected by the two-loop Barr-Zee diagrams
with the fermions loops and $W$ loops. Using the
well-known classical formulates \cite{mu2h13,mua2loop}, the main
contributions of two-loop Barr-Zee diagrams in the exact alignment
limit are given as
\bea
\label{mua2} \delta a_{\mu2}
&=&-\frac{\alpha m_\mu}{4\pi^3m_f}\sum_{\phi=h,H,A;f=t,b,\tau}
N_f^c~Q_f^2~ y_{\phi\mu\mu}~ y_{\phi ff}~ F_\phi(x_{f\phi})
\nonumber\\
&&+\frac{\alpha m_\mu}{8\pi^3v}\sum_{\phi=h} y_{\phi\mu\mu}~g_{\phi
WW} \left[3F_h\left(x_{W\phi}\right)
    +\frac{23}{4} F_A\left(x_{W\phi}\right)
     \right.\nonumber\\
    &&\left.
+\frac{3}{4} G\left(x_{W\phi}\right) +\frac{m_\phi^2}{2
m_W^2}\left\{
    F_h\left(x_{W\phi}\right)-F_A\left(x_{W\phi}\right)
    \right\}\right],
\eea
where $x_{f\phi}=m_{f}^2/m_\phi^2$, $x_{W\phi}=m_W^2/m_\phi^2$,
$g_{h WW}=1$ and
\begin{align}
  F_{\phi}(y)&=\frac{y}{2}\int_0^1 dx \frac{1-2x(1-x)}{x(1-x)-y}\log \frac{x(1-x)}{y}
~~({\rm for}~\phi=h,~H), \\
  F_\phi(y) &=\frac{y}{2}\int_0^1 dx \frac{1}{x(1-x)-y}\log \frac{x(1-x)}{y}~~
({\rm for}~\phi=A) , \\
  G(y)&=-\frac{y}{2}\int_0^1 dx \frac{1}{x(1-x)-y}\left[
    1-\frac{y}{x(1-x)-y}\log \frac{x(1-x)}{y}
    \right].
\end{align}
The difference between the SM value and the experimental value of muon $g-2$ is
\beq
\delta a_{\mu} =(26.2\pm8.5) \times 10^{-10}.
\eeq

\item[(3)] Lepton universality from the $\tau$ decays.
The current experimental results of
the charged lepton universality from $\tau$ decays are given by \cite{taudecay}
\beq
\frac{g_\mu}{g_e}=1.0018\pm0.0014,~~~\frac{g_\tau}{g_e}=1.0029\pm0.0015,
~~~\frac{g_{\tau}}{g_\mu}=1.0001\pm0.0014,
\label{leptonratio}\eeq
where the first two values are from the fit to
the leptonic decays of $\tau$, and third value is from the fit to
$\bar{\Gamma}(\tau\to e \nu\bar{\nu})/\bar{\Gamma}(\mu\to e
\nu\bar{\nu})$, and $\bar{\Gamma}(\tau\to h\nu)/\bar{\Gamma}(h\to
\mu\nu)$ with $h=K,\pi$ and $\bar{\Gamma}$ denoting the partial width
normalized to its SM value. The ratio $g_\tau/g_e$ favors
a positive correction to the SM value, which
will impose strong constraints on the 2HDM, as shown in \cite{mu2h10}.
Since only two of the ratios in Eq. (\ref{leptonratio}) are independent,
in principle we may take $g_\mu/g_e$ and $g_{\tau}/g_\mu$ to constrain the
model.

In this model, \beq\label{gmue}
(\frac{g_\mu}{g_e})^2=\bar{\Gamma}(\tau\to \mu
\nu\bar{\nu})/\bar{\Gamma}(\tau\to e
\nu\bar{\nu})\simeq1+\frac{z^2}{4}-\frac{2m_\mu}{m_\tau}z,\eeq
where $z=m_\mu m_\tau \kappa_\ell^2/m^2_{H^{\pm}}$.
The corrections are from the tree-level diagrams mediated
by the charged Higgs. Since the one-loop effect applies equally to both tau decays, it
does not give the correction to $\frac{g_\mu}{g_e}$.

Ignoring the electron mass,
$g_{\tau}/g_\mu$ is only corrected by the one-loop diagram mediated by the charged Higgs.
The corrections to $g_\tau/g_\mu$ is given by \cite{mu2h10}
\beq
(\frac{g_\tau}{g_\mu})^2=\bar{\Gamma}(\tau\to e
\nu\bar{\nu})/\bar{\Gamma}(\mu\to e \nu\bar{\nu})\simeq1+2\delta
g,
\eeq
where
\beq
\delta g=\frac{1}{16\pi^2}(\frac{m_\tau}{v}\kappa_\ell)^2
\left[1+\frac{m_{H^\pm}^2+m_A^2}{4(m_{H^\pm}^2-m_A^2)}
\log\frac{m_A^2}{m_{H^\pm}^2}+\frac{m_{H^\pm}^2+m_H^2}{4(m_{H^\pm}^2-m_H^2)}
\log\frac{m_H^2}{m_{H^\pm}^2}\right].
\eeq

\item[(4)] The measurements of $R(D^{(*)})$.
The new four-fermion operators can be generated by exchanging the charged Higgs:
\begin{align}
  {\mathcal O}^{\,q}_{\rm SRL}&=(\bar q P_R b)( \bar\tau P_L \nu_\tau),\\
  {\mathcal O}^{\,q}_{\rm SLL}&=(\bar q P_L b)( \bar\tau P_L \nu_\tau)\,.
\end{align}
The corresponding tree-level Wilson coefficients are given by
\begin{eqnarray}
\label{eq:WC2}
  C_{\rm SLL}^{\,c}=\frac{2\sqrt{m_tm_c}m_\tau}{M_{H^{\pm}}^2v^2}V_{tb} \lambda_{ct} \kappa_{\ell}\,,~~~
    C_{\rm SRL}^{\,c}=-\frac{2m_bm_\tau}{M_{H^{\pm}}^2v^2}V_{cb} \kappa_{d} \kappa_{\ell}\,,
\end{eqnarray}
which can give the contributions to $R(D^{(*)})$ \cite{Sakaki:2012ft,rd1,rd2,rd3}
\begin{align}
\label{eq:RDexpr}
  R(D)  &=R_{\rm SM}(D)\left (1+1.5{\rm Re}\left [\frac{C^{\,c}_{\rm SRL}+C^{\,c}_{\rm SLL}}{C_{\rm VLL}^{\,c,\,\rm SM}} \right] + 1.0 \left\lvert \frac{C^{\,c}_{\rm SRL}+C^{\,c}_{\rm SLL}}{C_{\rm VLL}^{\,c,\,\rm SM}} \right\rvert^2 \right),\nonumber\\
  R(D^*)  &=R_{\rm SM}(D^*)\left (1+0.12{\rm Re}\left [\frac{C^{\,c}_{\rm SRL}-C^{\,c}_{\rm SLL}}{C_{\rm VLL}^{\,c,\,\rm SM}} \right] + 0.05 \left\lvert \frac{C^{\,c}_{\rm SRL}-C^{\,c}_{\rm SLL}}{C_{\rm VLL}^{\,c,\,\rm SM}} \right\rvert^2 \right).
\end{align}
Here $C_{\rm VLL}^{\,c,\,\rm SM}$ is the Wilson coefficient in the SM
\begin{equation}
\label{eq:WC1}
 C_{\rm VLL}^{\,c,\,\rm SM}=\frac{4G_F V_{cb}}{\sqrt 2}\,.
\end{equation}

\item[(5)] $B$-meson decays. The non-diagonal matrix element $X_{tc}$ can give additional contributions to
the couplings of top quark and charged Higgs, which will correct $\Delta m_{B_s}$, $\Delta m_{B_d}$ and $B\to X_s\gamma$
at the one-loop level:
 \begin{align} \label{topch}
 & H^+ \bar{t}s: - \frac{\sqrt{2}m_t}{v}V_{ts} \left(\kappa_u+\sqrt{\frac{m_c}{m_t}}\frac{V_{cs}}{V_{ts}}\lambda_{ct}\right) P_L
                 +  \frac{\sqrt{2}m_s}{v}V_{ts} \kappa_d P_R,\\
 & H^+ \bar{t}b: - \frac{\sqrt{2}m_t}{v}V_{tb} \left(\kappa_u+\sqrt{\frac{m_c}{m_t}}\frac{V_{cb}}{V_{tb}}\lambda_{ct}\right) P_L
                 +  \frac{\sqrt{2}m_b}{v}V_{tb} \kappa_d P_R.
 \end{align}
The $\Delta m_{B_s}$, $\Delta m_{B_d}$ and $B\to X_s\gamma$
are respectively calculated using the
formulas in \cite{deltmq,bsr1,bsr2}.

\item[(6)] Higgs search experiments:
\begin{itemize}
\item[(i)] The non-observation of additional Higgs bosons. We employ
$\textsf{HiggsBounds-4.3.1}$ \cite{hb1,hb2} to implement the exclusion
constraints from the neutral and charged Higgs searches at LEP,
Tevatron and LHC at 95\% confidence level.

Very recently, ATLAS reported the searches for a heavy charged
Higgs in the single top quark associated production at the 13 TeV
LHC with an integrated luminosity of 14.7 fb$^{-1}$ for $H^\pm\to
\tau\nu$ \cite{hptav} and 13.2 fb$^{-1}$ for $H^\pm\to tb$
\cite{hptb}. The upper bounds of production cross section times
Br$(H^\pm \to \tau\nu)$ are in the range of 2.0 to 0.008 pb for
$m_{H^{\pm}}=$ 200-2000 GeV. The upper bounds of production cross
section times Br$(H^\pm \to tb)$ are in the range of 1.37 and 0.18
pb for $m_{H^{\pm}}=$ 300-1000 GeV. In the model, the top quark FCNC
of the charged Higgs can give additional contributions to the
production of a charged Higgs boson in association with a top quark.
Although the coupling of the charged Higgs and tau lepton is sizably
enhanced, the decay $H^\pm \to A W^\pm$ is still an important mode.

\item[(ii)] The global fit to the 125 GeV Higgs signal data.
In the exact alignment limit, the SM-like Higgs couplings to the SM
particles at tree-level are the same as the SM, which is favored by
the 125 GeV Higgs signal data. For $m_A< 62.5$ GeV, the mode $h\to
AA$ can open and enhance the total width of $h$ sizably, which will
be strongly constrained by the 125 GeV Higgs data. We perform the
$\chi^2$ calculation for the signal strengths in the
$\mu_{ggF+tth}(Y)$ and $\mu_{VBF+Vh}(Y)$ with $Y$ denoting the decay
mode $\gamma\gamma$, $ZZ$, $WW$, $\tau\bar{\tau}$ and $b\bar{b}$,
 \begin{eqnarray} \label{eq:ellipse}
  \chi^2(Y) =\left( \begin{array}{c}
        \mu_{ggH+ttH}(Y) - \widehat{\mu}_{ggH+ttH}(Y)\\
        \mu_{VBF+VH}(Y) - \widehat{\mu}_{VBF+VH}(Y)
                 \end{array} \right)^T
                 \left(\begin{array}{c c}
                        a_Y & b_Y \\
                        b_Y & c_Y
                 \end{array}\right) \nonumber\\
\times
                  \left( \begin{array}{c}
        \mu_{ggH+ttH}(Y) - \widehat{\mu}_{ggH+ttH}(Y)\\
        \mu_{VBF+VH}(Y) - \widehat{\mu}_{VBF+VH}(Y)
                 \end{array} \right) \,,
 \end{eqnarray}
where $\widehat{\mu}_{ggH+ttH}(Y)$ and $\widehat{\mu}_{VBF+VH}(Y)$
are the data best-fit values and $a_Y$, $b_Y$ and $c_Y$ are the
parameters of the ellipse. These parameters are given by the
combined ATLAS and CMS experiments \cite{160602266}. We pay
particular attention to the surviving samples with
$\chi^2-\chi^2_{\rm min} \leq 6.18$, where $\chi^2_{\rm min}$
denotes the minimum of $\chi^2$. These samples correspond to the
95.4\% confidence level region in any two-dimension plane of the
model parameters when explaining the Higgs data (corresponding to
the $2\sigma$ range).
\end{itemize}

\item[(7)] Observables of the top quark:
\begin{itemize}
\item[(i)] The total width of the top quark. There is no decay mode $t\to hc$ in the exact alignment limit.
For $m_A<m_t$, the mode $t\to Ac$ will open and enhance the total width of the top quark.
The measurement value of the total top width is $\Gamma_t^{exp}=1.36 \pm 0.02^{+0.14}_{-0.11}$  GeV from
the CMS collaboration \cite{topwid}.

\item[(ii)] Same-sign top pair production at the LHC. The same-sign top pair can be produced at the LHC
via the $c~c\to t~ t$ process with the $t$-channel exchange of $A$
and $H$. From the searches for the same-sign dileptons and b-jets at
the 8 TeV LHC with an integrated luminosity of 20.3 fb$^{-1}$
\cite{tt}, ATLAS gave an upper bound of 62 fb.
\end{itemize}
\end{itemize}

Note that the observables of the top quark, $B$-meson decays and the searches for the heavy charged
Higgs at the 13 TeV LHC are all sensitive to the FCNC of top quark, namely the parameter $\lambda_{ct}$.
For convenience, we use "Top-FCNC-Constraints" to denote these constraints in the following sections.
We perform $\textsf{MG5@NLO}$ \cite{mg5} to calculate
the cross section of $pp\to tt$ at the 8 TeV LHC and $\sigma(pp\to tH^{\pm})\times Br(H^{\pm}\to \tau\nu,~ tb)$
at the 13 TeV LHC. In fact, our calculations show that the searches for the same-sign top pair production
at the 8 TeV LHC and the charged Higgs at the 13 TeV LHC can hardly give further constraints on the model
after imposed the constraints from the $B$-meson decays, top width, muon $g-2$, $R(D^{(*)})$, $\tau$ decays,
precision electroweak data and theoretical constraints.

\section{Results and discussions}
\subsection{Explanation for $R(D^{(*)})$ and muon $g-2$}
\begin{figure}[tb]
 \epsfig{file=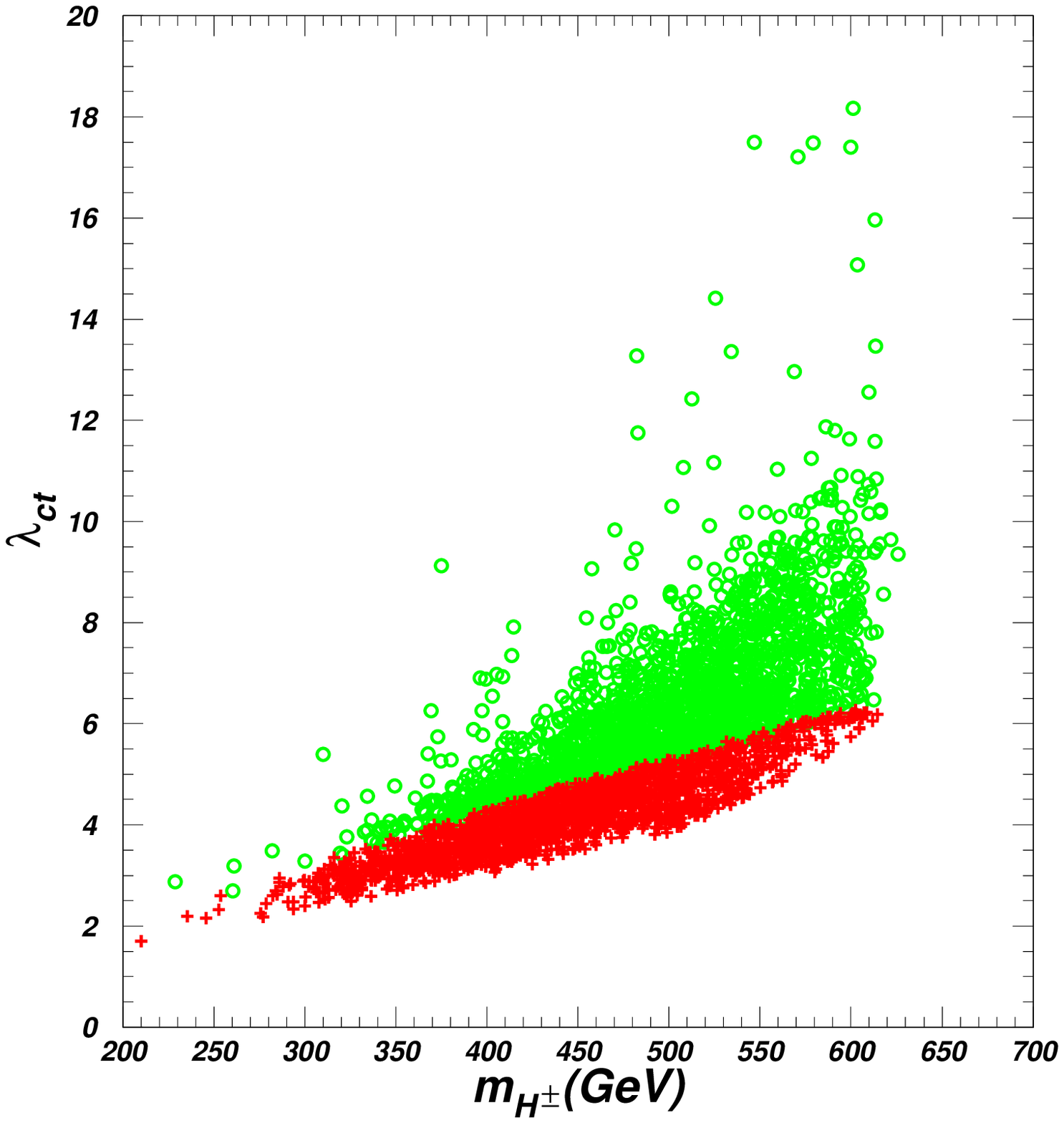,height=5.71cm}
  \epsfig{file=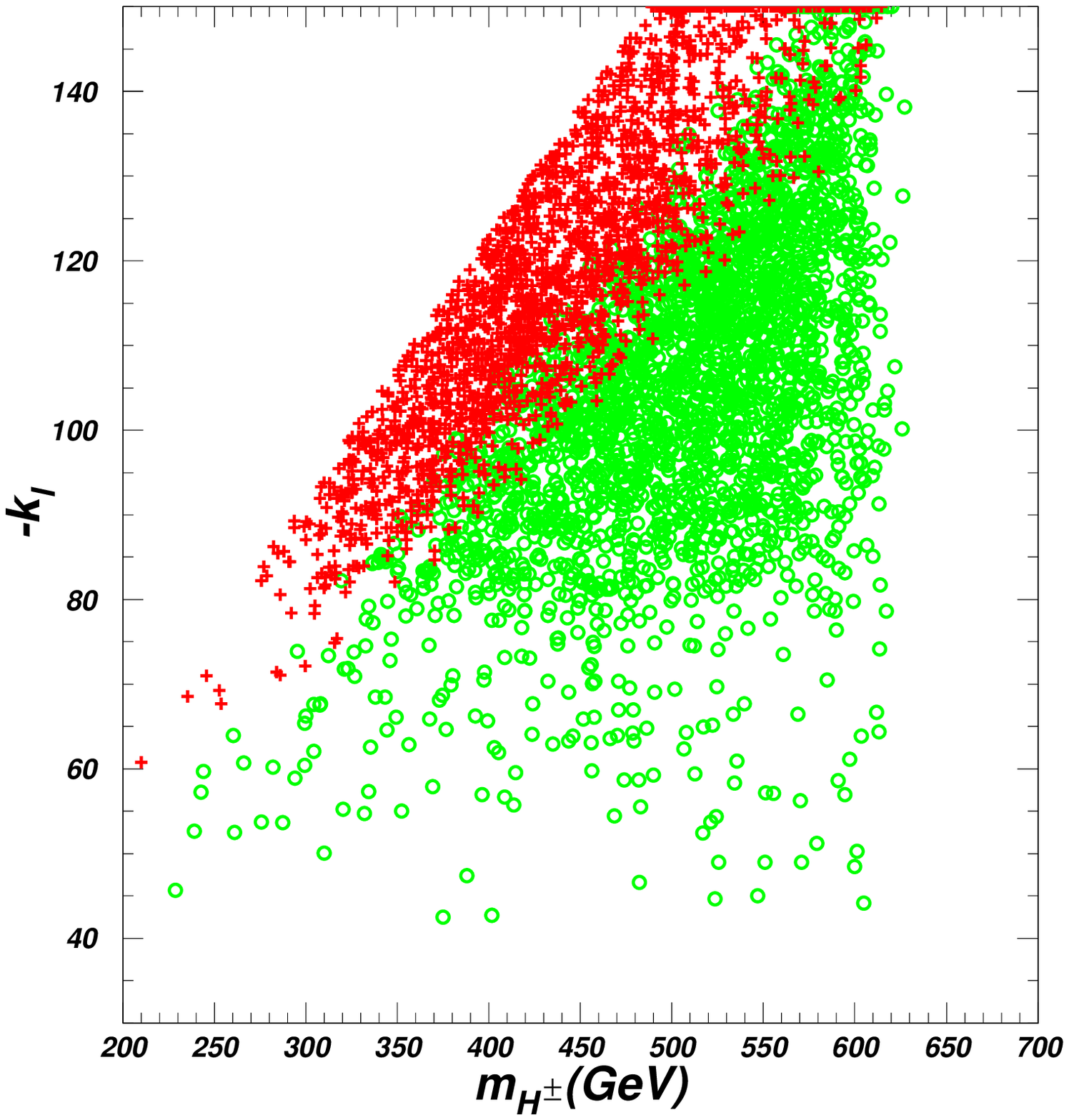,height=5.71cm}
 \epsfig{file=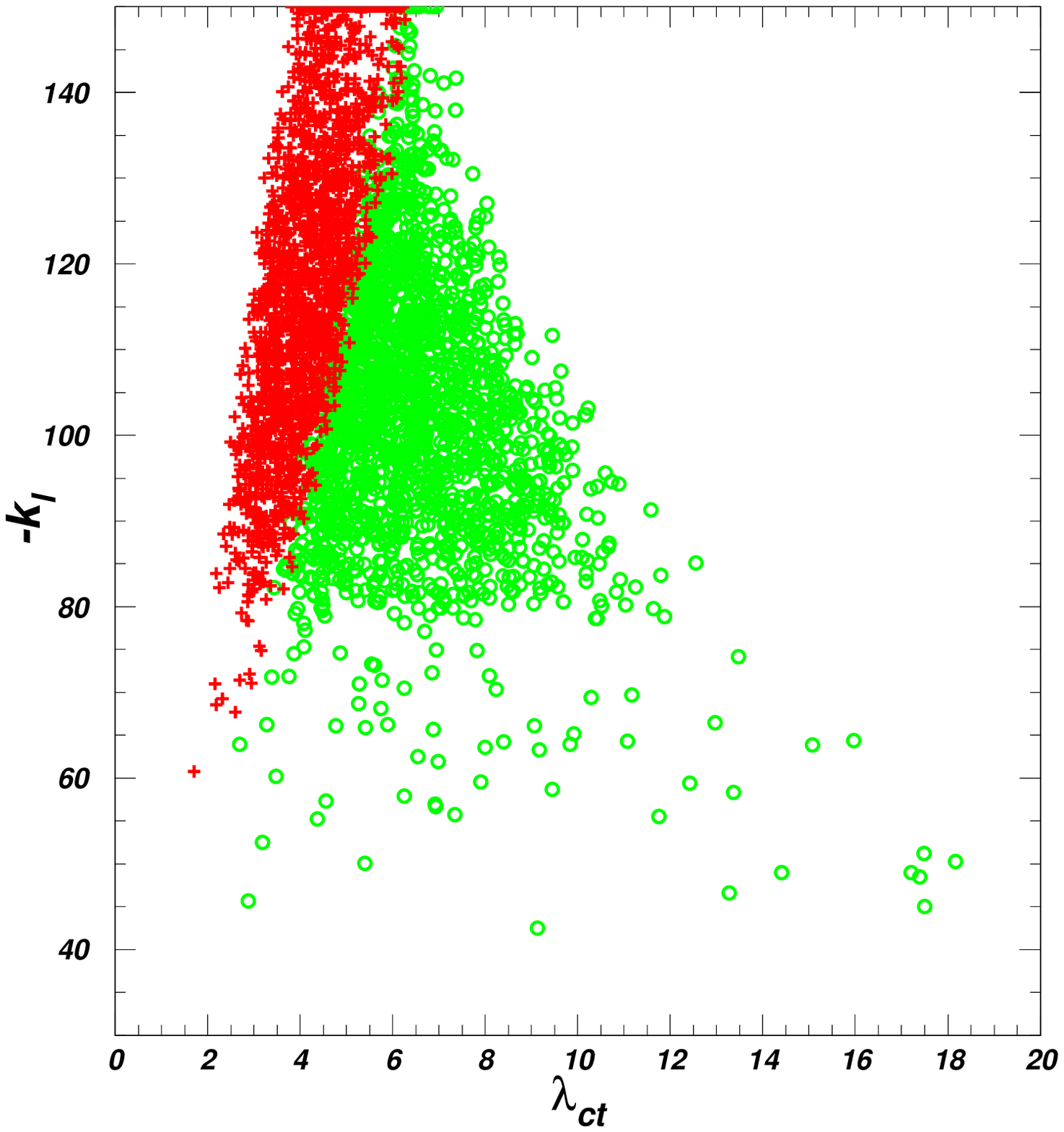,height=5.71cm}
\vspace{-1.0cm} \caption{The surviving samples projected on the
planes of $\lambda_{ct}$ versus $m_{H^\pm}$, $\kappa_\ell$
versus $m_{H^\pm}$ and $\kappa_\ell$
versus $\lambda_{ct}$. All the points are allowed by the constraints of the muon $g-2$, $R(D^{(*)})$,
the theoretical constraints, the precision electroweak data, the $\tau$ decays, the exclusion limits
of Higgs bosons and the 125 GeV Higgs data.
The circles (green) and pluses (red) are respectively excluded and allowed by
the "Top-FCNC-Constraints".} \label{figrd}
\end{figure}

In Fig. \ref{figrd}, we project the surviving samples on the planes
of $\lambda_{ct}$ versus $m_{H^{\pm}}$, $\kappa_\ell$ versus
$m_{H^{\pm}}$ and $\kappa_\ell$ versus $\lambda_{ct}$. Without the
"Top-FCNC-Constraints", the $\lambda_{ct}$ and $|\kappa_\ell|$ increase
with the charged Higgs mass, and $|\kappa_\ell|$ tends to decrease with the
increasing of $\lambda_{ct}$. These features are mainly determined
by $R(D^{(*)})$ since the product $\lambda_{ct}\kappa_\ell /
m^2_{H^\pm}$ in the Wilson coefficient $C_{SLL}^{c}$ can affect
$R(D^{(*)})$ (see Eq. (\ref{eq:WC2}) and Eq. (\ref{eq:RDexpr})).
 In addition, the observables of the lepton universality from
$\tau$ decays favor $|\kappa_\ell|$ to increase with the charged Higgs mass mainly due to
the factor $\kappa^2_\ell/m^2_{H^\pm}$ in the correction terms of $g_\mu/g_e$ (see Eq. (\ref{gmue})).

After imposing the "Top-FCNC-Constraints", a large part of the parameter space is excluded,
and $\lambda_{ct}$ and $m_{H^\pm}$ are directly constrained. For
a given $m_{H^\pm}$, $\lambda_{ct}$ will be imposed an upper bound
by the "Top-FCNC-Constraints" and a lower
bound by $R(D^{(*)})$. Once $m_{H^\pm}$ and the upper bound of
$\lambda_{ct}$ are given, $R(D^{(*)})$ will impose a lower bound on
$|\kappa_\ell|$. In addition, the lepton universality from $\tau$
decays will give an upper bound on $|\kappa_\ell|$ for a given
$m_{H^\pm}$. For example, 3.0 $<\lambda_{ct}<$ 4.5 and 90
$<|\kappa_\ell|<$ 125 for $m_{H^\pm}=400$ GeV. After imposing "Top-FCNC-Constraints",
$\lambda_{ct}$ and $|\kappa_\ell|$ increase with the charged Higgs
mass, and $|\kappa_\ell|$ tends to increase with $\lambda_{ct}$. For
200 GeV $<m_{H^\pm}<620$ GeV, $\lambda_{ct}$ and $|\kappa_\ell|$ are
respectively required to be in the ranges of 1.5 $\sim$ 6.5 and 60
$\sim$ 150.

\begin{figure}[tb]
  \epsfig{file=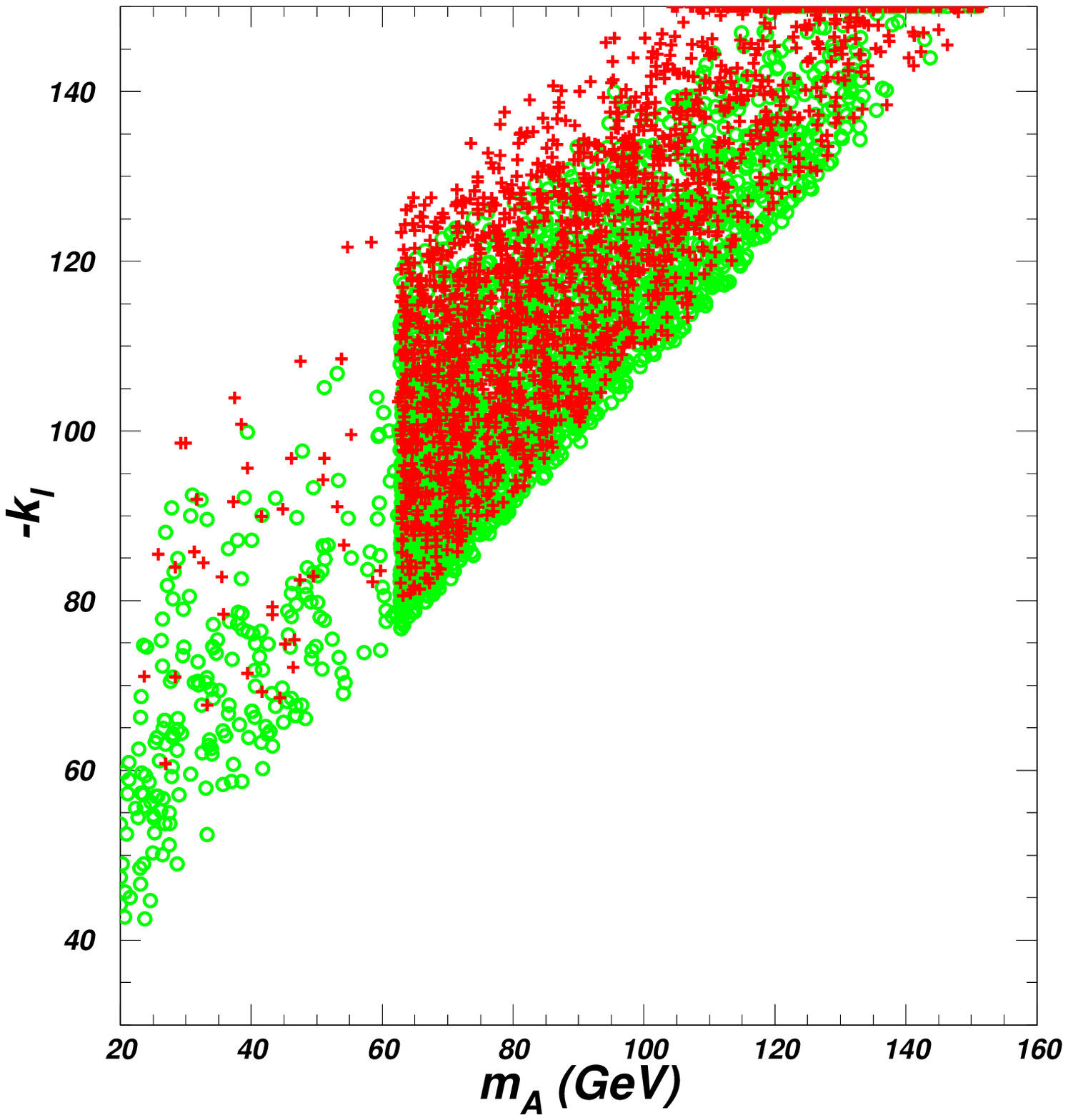,height=7.3cm}
  \epsfig{file=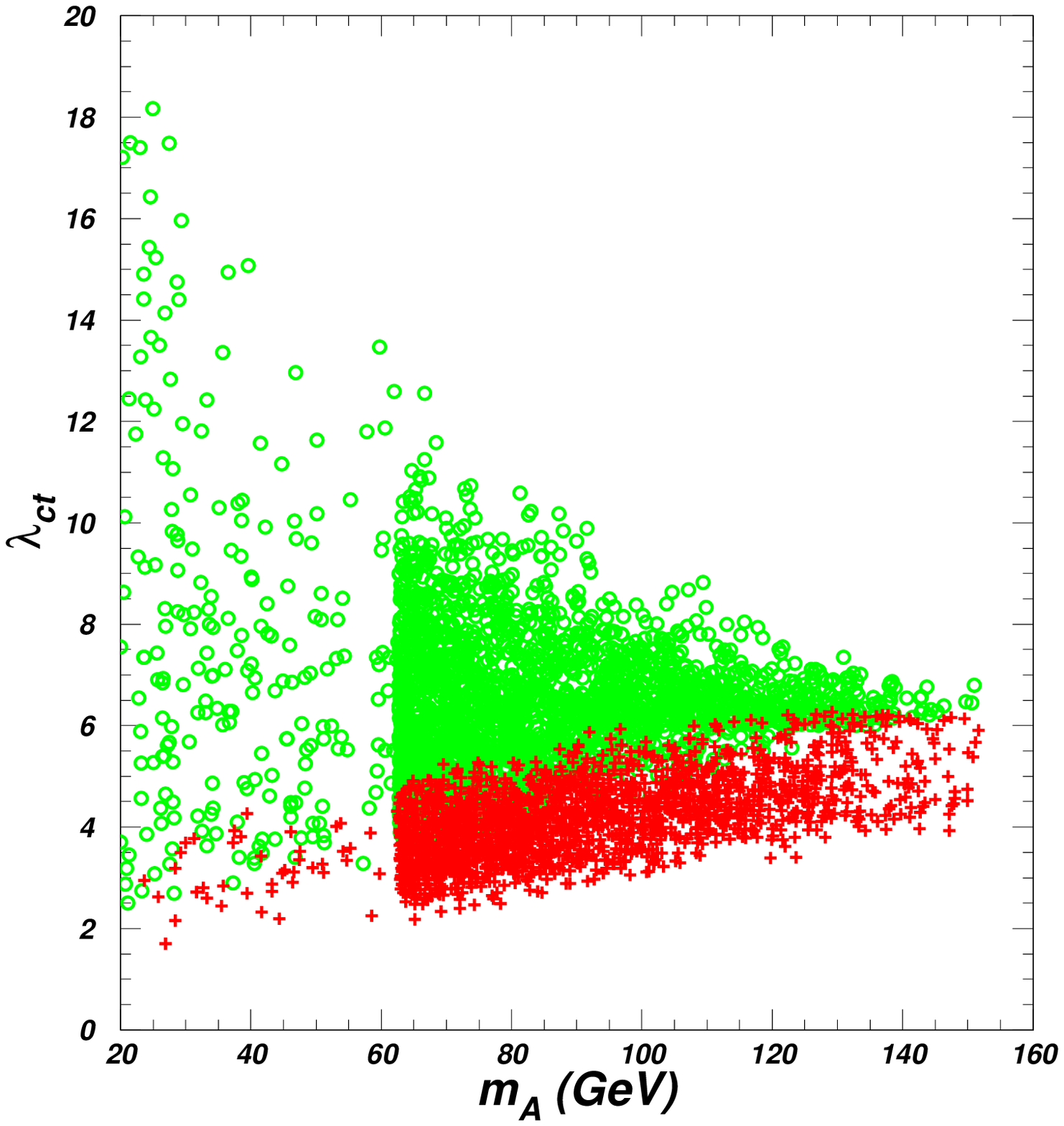,height=7.3cm}
\vspace{-0.5cm} \caption{Same as Fig. \ref{figrd}, but showing $\kappa_\ell$
versus $m_A$ and $\lambda_{ct}$ versus $m_A$.} \label{figmug2}
\end{figure}

In Fig. \ref{figmug2}, we project the surviving samples on the
planes of $\kappa_{\ell}$ versus $m_A$ and $\lambda_{ct}$ versus
$m_A$. The left panel shows that $|\kappa_\ell|$ is sensitive to
$m_A$, and increases with $m_A$. These features are mainly
determined by the muon $g-2$ which is given the dominantly positive
contributions by the pseudoscalar via the two-loop Barr-Zee
diagrams. For $m_A>150$ GeV, $|\kappa_\ell|$ is required to be
larger than 150, which potentially leads to a problem with the
perturbativity of the lepton Yukawa coupling. For $m_A=20$ GeV,
$|\kappa_\ell|$ is required to be larger than 60 after imposing
"Top-FCNC-Constraints". Compared to the region of $m_A > 62.5$ GeV,
there are relatively few surviving samples in the region of $m_A <
62.5$ GeV. For $m_A$ smaller than the half of the SM-like Higgs
mass, the decay mode $h\to AA$ will open and enhance the total width
of the SM-like Higgs. Therefore, the data of the 125 GeV Higgs can
give strong constraints on the parameter space.

The right panel of Fig. \ref{figmug2} shows that the surviving samples favor a large $\lambda_{ct}$ for
a large $m_A$ to after imposing the "Top-FCNC-Constraints". Since the contributions of $A$ and $H$
to muon $g-2$ canceled, a large mass splitting of $A$ and $H$ is required to
explain the muon $g-2$. In addition, the precision electroweak data
favor a small mass splitting of $H$ and $H^{\pm}$. Therefore, the muon $g-2$ favors
a large $m_{H^\pm}$ for a large $m_A$, and further a large $m_{H^\pm}$ tends to require a
large $\lambda_{ct}$ due to the $R(D^{(*)})$ excess and "Top-FCNC-Constraints".

Note that flipping the signs of $\lambda_{ct}$ and $\kappa_\ell$
does not change the results in this paper. As seen from section III,
the muon $g-2$ and the observables of lepton universality do not
depend on the sign of $\kappa_\ell$. $R(D^{(*)})$ depends on the
sign of the product of $\lambda_{ct}\kappa_\ell$. When flipping the
signs of $\lambda_{ct}$ and $\kappa_\ell$, the sign of the charged
Higgs coupling of top quark will be flipped and the absolute value
remains unchanged due to $\kappa_u=\kappa_d=-\kappa_\ell$, which
does not change the results of $B$-meson decays.

\subsection{Simulation on $pp\to t\bar{t}\to WbAc\to jj b c \tau\bar{\tau}$}
As seen from the preceding section, after imposing the relevant
theoretical and experimental constraints, the muon $g-2$ and
$R(D^{(*)})$ excesses can be simultaneously explained in the parameter space:
\begin{align}
\label{spaceyes}
&20~{\rm GeV} <m_A < 150~ {\rm GeV},~~ 200~ {\rm GeV}< m_H~(m_{H^{\pm}})< 620~ {\rm GeV},\nonumber\\
& -150 <\kappa_\ell< -60,~~\kappa_u=\kappa_d=-1/\kappa_\ell,~~1.5<\lambda_{ct}<6.5.
\end{align}
In such a parameter space, the pseudoscalar can be produced via the QCD process $pp\to t\bar{t}$
followed by the decay $t\to A c$, and then dominantly decays into $\tau\bar{\tau}$.

In the parameter space shown in Eq. (\ref{spaceyes}), the decay
modes $A\to HZ$, $A\to H^\pm W^\mp$ and $A\to hZ$ are kinematically
forbidden, and $A\to hZ$ is also absent in the exact alignment
limit. The pseudoscalar will dominantly decay into $\tau\bar{\tau}$,
$Br(A\to \tau\bar{\tau})\simeq99.65\%$ and $Br(A\to
\mu\bar{\mu})\simeq0.35\%$, which are not sensitive to
$|\kappa_\ell|$ in the range of 60 to 150. Therefore, here the cross
section $pp\to t\bar{t}$ times $Br(t\to Ac\to \tau\bar{\tau}c)$ is
only sensitive to $m_A$ and $\lambda_{ct}$. Besides, since the cross
sections of $pp\to A$ and $pp\to At\bar{t}$ are sizably suppressed
by $\kappa^2_u$, the $pp\to t\bar{t}\to WbAc$ production process
becomes more important.

Now we perform detailed simulations on the signal and backgrounds at the 13 TeV LHC.
We consider the top quark pair production where one decays to $(W\to j j )+b$ and
the other decays to $(A\to\tau\bar{\tau})+c$. The major SM background processes to
this signal are $t\bar{t}$, $tW+$jets, $tZ+$jets and $Zb\bar{b}$+jets.
Other backgrounds, such as the multi-jets,
can be significantly reduced by requiring one b-tagged jet, two $\tau$-tagged jets.

The model file of the 2HDM is generated by $\textsf{FeynRules}$ \cite{feyrule}.
Both the signal and background processes are generated with $\textsf{MG5@NLO}$ \cite{mg5},
using $\textsf{PYTHIA}$ for showering and hadronization \cite{pythia},
$\textsf{TAUOLA}$ for $\tau$ lepton decay \cite{TAUOLA}.
The fast simulations of the detector and trigger are performed by $\textsf{Delphes3.3.0}$ \cite{delphes},
including $\textsf{Fastjet3}$ for jet clustering \cite{Fastjet3}.

We identify the lepton candidates by requiring them to have $p_T>15$
GeV and $|\eta|<2.5$. The anti-kt algorithm is employed to
reconstruct the jets with a radius parameter $R = 0.4$ \cite{kt},
and the jets are required to have $p_T>20$ GeV and $|\eta|<2.5$. We
assume an average b-tagging efficiency of 70\% for real b-jets, with
misidentification efficiency of 10\%, 4\% and 0.2\% for c-jets,
$\tau$-jets and jets initiated by light quarks or gluons
respectively. We use the medium hadronic $\tau$ identification
criteria with an efficiency of about 55\% \cite{medium}. In order to
suppress multi-jet backgrounds, we also require that the jets
separated by $R < 0.2$ from the $\tau$-tagged jet are removed.

According to the signal topology, we consider a final state of more than six jets
including exactly two tau-tagged jets, one or two b-tagged jets,
with missing transverse momentum $E_T^{miss}> 20$ GeV.
Events with electrons or muons are vetoed.
For event selection we require that the $E_T^{miss}$
centrality $C_{miss}$ is greater than zero \cite{1509.08149}:
\begin{equation}
\begin{aligned}
  C_{miss} = \frac{x+y}{\sqrt{x^2+y^2}}, ~x=\frac{sin(\phi_{miss}-\phi_{\tau1})}{sin(\phi_{\tau2}-\phi_{\tau1})},y=\frac{sin(\phi_{\tau2}-\phi_{miss})}{sin(\phi_{\tau2}-\phi_{\tau1})},
\end{aligned}
\end{equation}
where $\phi_{miss}$ is the azimuthal angle of $E_T^{miss}$, and $\phi_{\tau1,2}$
are the azimuthal angle of the two tau-tagged jets in the transverse plane.

\begin{figure}[tb]
   \epsfig{file=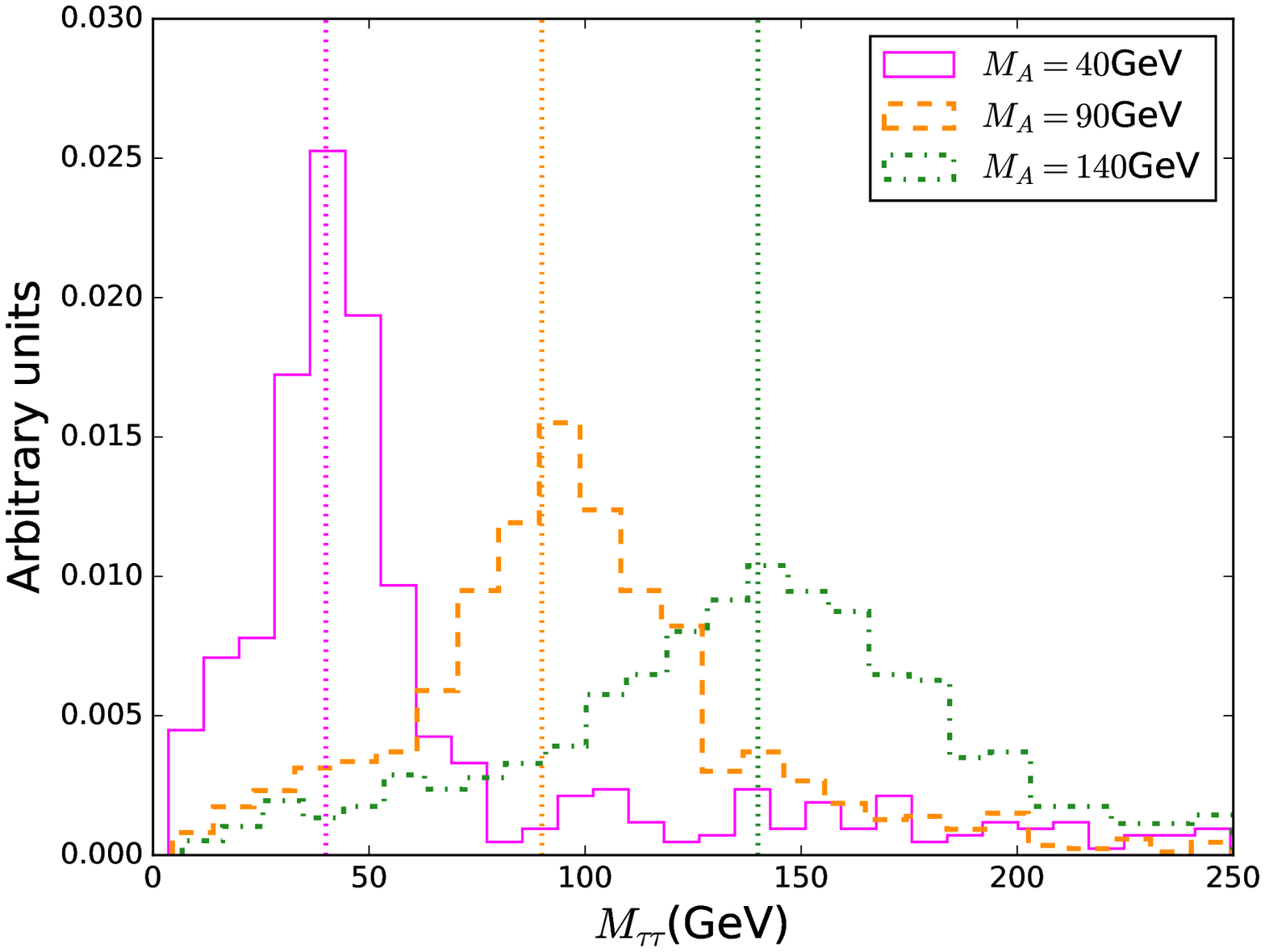,height=6cm}
  \epsfig{file=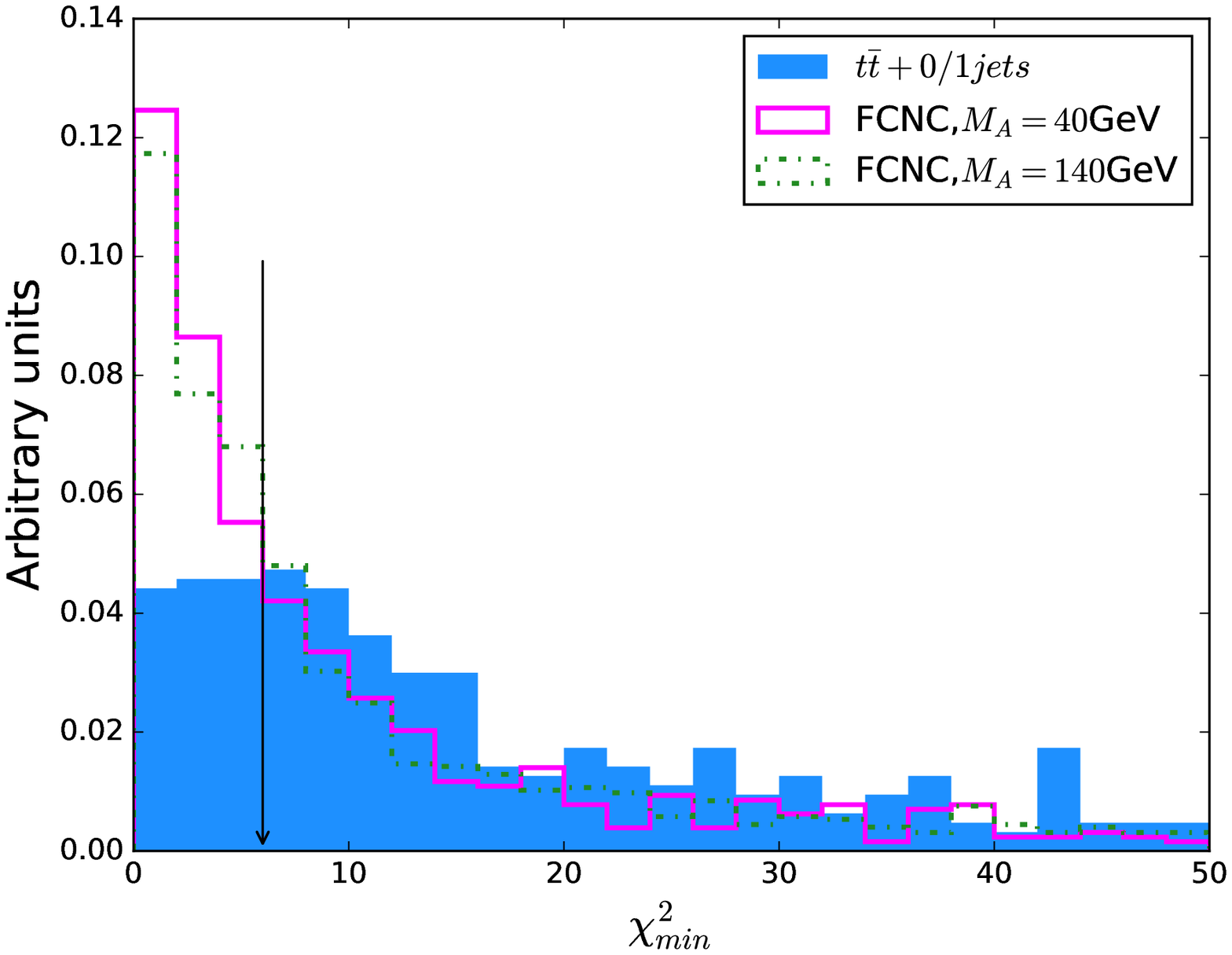,height=6cm}
\vspace{-.3cm}
   \caption{The left panel is the distributions of the reconstructed mass $M_{\tau\tau}$
for the signal events with $M_A=40$ GeV (solid line),
$M_A=90$ GeV (dashed line) and $M_A=140$ GeV (dash-dotted line).
The right panel is the distributions of $\chi^2_{min}$ for
the signal events with $M_A=40$ GeV (solid line) and
$M_A=140$ GeV (dash-dotted line) and for the main background
$t\bar{t}+0/1$ jet events (bins). The arrow represents the cut
used in the event selection. All distributions are normalized to the
unit area.} \label{Fig:distribution}
 \end{figure}

To suppress the backgrounds, we reconstruct the kinematics of the top quarks
 from the corresponding decay particles. Firstly, because of the presence of
 the neutrino in $\tau$ hadronic decay, we use the collinear approximation
technique to determine the 4-momenta of neutrino \cite{1012.4686},
which is based on two assumptions: the neutrinos from each $\tau$ are collinear
with the corresponding visible $\tau$ decay products and $E_T^{miss}$ is
only due to neutrinos. For our signal events, there is no other $E_T^{miss}$
contribution and the $\tau$ leptons are from the cascade decay of the top
quark which can be boosted depending on $m_A$. The invisible momentum of
neutrino in each $\tau$ decay is determined by
\begin{equation}
\begin{aligned}
  & E_T^{miss} {\rm cos} \phi_{miss} = p_{mis1} {\rm sin}\theta_{vis1} {\rm cos}\phi_{vis1} + p_{mis2} {\rm sin}\theta_{vis2} {\rm cos}\phi_{vis2}, \\
  & E_T^{miss} {\rm sin} \phi_{miss} = p_{mis1} {\rm sin}\theta_{vis1} {\rm sin}\phi_{vis1} + p_{mis2} {\rm sin}\theta_{vis2} {\rm sin}\phi_{vis2}.
\end{aligned}
\end{equation}
where $\phi_{miss}$ is the azimuthal angle of $E_T^{miss}$, $\theta_{vis1,2}$ and $\phi_{vis1,2}$
are the polar and azimuthal angles of the $\tau$ jets, and $p_{mis1}$ and $p_{mis2}$ are the invisible
momentum of $\tau$ decay.
Then one can obtain the invariant mass of the two
 $\tau$ leptons $M_{\tau\tau}$ and compare with the mass of the pseudoscalar $A$.
In the left panel of Fig. \ref{Fig:distribution}, we show some
examples for the distributions of $M_{\tau\tau}$ for the signal
events with $M_A=40$ GeV, $90$ GeV and $140$ GeV. One can see that
this technique is more effective for a small $m_A$.

Since the mass of the pseudoscalar $A$ is unknown, we use the reconstruct mass of
top quark $m_{j_c\tau\tau}$ instead of $M_{\tau\tau}$.
Together with the reconstructed masses of the SM decay of top quark
and the W boson, $m_{j_bj_1j_2}$ and $m_{j_1j_2}$, we define a
$\chi^2$ function as
\begin{equation}
\begin{aligned}
  \chi^2=\frac{(m_{j_c\tau\tau}-m_{FCNC})^2}{\sigma_{FCNC}^2}+\frac{(m_{j_bj_1j_2}-m_t)^2}{\sigma_{SM}^2}+\frac{(m_{j_1j_2}-m_W)^2}{\sigma_{W}^2},
\end{aligned}
\end{equation}
where $m_{FCNC}=153$ GeV, $m_t=173$ GeV, $\sigma_{FCNC}=20$ GeV, $\sigma_{SM}=20$ GeV, $m_W=82$ GeV and $\sigma_{W}=15$
 GeV taken from \cite{1508.05796,1509.08149}. As we keep the events containing two b-tagged jets
 in which one of them is misidentified for the charm quark from the $t\to Ac$,
the assignment of each jet to the reconstruct masses is dependent on the number
of b-tagged jets. For the events involving two b-tagged jets, any b-tagged
jet can be assigned to $j_c$ and $j_b$, while $j_1$ and $j_2$ correspond
to the leading two light-flavor jets. For events containing one b-tagged jet,
the b-tagged jet is referred to $j_b$,  $j_c$ is chosen from the leading three
light-flavor jets, and the other two jets in the leading three light-flavor jets
 act as $j_1$, $j_2$. From all the combinations, the one with the
minimum $\chi_{min}^2$ is chosen, and then we require $\chi_{min}^2<6$
according to the distributions of $\chi^2_{min}$ for signal and
background events as shown in Fig \ref{Fig:distribution}.
\begin{figure}[tb]
  \epsfig{file=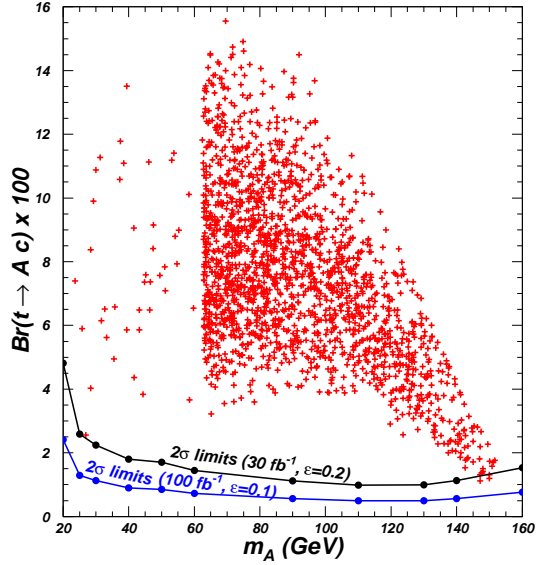,height=7.5cm}
\vspace{-0.6cm}
  \caption{The $2\sigma$ upper limits on Br($t\to A c$) as functions of $m_A$
for a data set of 30 fb$^{-1}$ and 100 fb$^{-1}$ at the 13 TeV LHC.
All the samples satisfy the relevant constraints and
can simultaneously explain the muon g-2 and $R(D^{(*)})$ excesses.} \label{Fig:significance}
\end{figure}

After imposing the above selection conditions, the cross section of $t\bar{t}$
at the 13 TeV LHC is reduced to 61.3 fb, while $Z+b\bar{b}/c\bar{c}$ and $tZ$ are reduced to 0.12 fb
and 0.03 fb, respectively.
Then we calculate the signal significance with simplified definition,
\begin{equation}
\begin{aligned}
  \mathcal{S} = \frac{n_s}{\sqrt{n_b+(\varepsilon n_b)^2}},
\end{aligned}
\end{equation}
where $n_s$ and $n_b$ are the expected numbers of the signal and
background event, and $\varepsilon$ is the relative systematic
uncertainty which we conservatively take 20\% and 10\% for a data
set of 30 fb$^{-1}$ and 100 fb$^{-1}$ at the 13 TeV LHC in our
analysis. In Fig. \ref{Fig:significance} we show the results in
plane of $m_A$ and Br($t\to A c$). All the samples in Fig.
\ref{Fig:significance} satisfy the relevant constraints and can
explain the muon $g-2$ and $R(D^{(*)})$ excesses simultaneously.
Depending on $m_A$, the branching ratio of $t\to Ac$ is required to
above 1\% and below 16\%. The $2\sigma$ upper limits from a data set
of 30 fb$^{-1}$ at the 13 TeV LHC can exclude almost all the
samples, and a few samples can survive only when $m_A$ approaches to
150 GeV or 20 GeV. However, the $2\sigma$ upper limits from a data
set of 100 fb$^{-1}$ at 13 TeV LHC can exclude all the samples.

\section{Conclusion}
In the framework of a two-Higgs-doublet model with top quark
FCNC couplings, we examined the excesses of $R(D^{(*)})$ and muon $g-2$ by
imposing the relevant theoretical and experimental constraints
from the precision electroweak data, $B$-meson decays, $\tau$ decays,
the observables of top quark and Higgs searches. In this model the coupling
$\kappa_\ell$ can simultaneously affect $R(D^{(*)})$, the muon $g-2$ and the
lepton universality from $\tau$ decays, and thus these three observables
have a strong correlation.

We found that the $R(D^{(*)})$ and muon $g-2$ excesses can be simultaneously
explained in the parameter space allowed by the relevant constraints.
In such a parameter space, the pseudoscalar is between 20 GeV and 150 GeV
so that it can be produced from the top quark FCNC decay
$t\to A c$ and then dominantly decays into $\tau\bar{\tau}$.
We performed a detailed simulation on the signal $pp\to t\bar{t}\to Wb Ac\to
jjbc\tau\bar{\tau}$ and the corresponding backgrounds, and found that
the $2\sigma$ upper limits from a data
set of 30 (100) fb$^{-1}$ at the 13 TeV LHC can mostly (totally) exclude
such a parameter space.

\section*{Acknowledgment}
We thank Xin-Qiang Li, Ying Li and and Zhi-Tian Zou for discussions.
 This work is supported by the National Natural Science Foundation
of China under grant Nos. 11575152, 11275245, 11135003, by the CAS
Center for Excellence in Particle Physics (CCEPP) and by the CAS Key
Research Program of Frontier Sciences.

\end{document}